\documentclass{article}
\usepackage{spconf,amsmath,graphicx}
\usepackage{booktabs}
\usepackage{amssymb}
\usepackage{mathtools, nccmath}
\usepackage{stfloats}
\usepackage{threeparttable}

\usepackage[english]{babel}
\usepackage{blindtext}

\newcommand{\norm}[1]{\left\lVert#1\right\rVert}

\title{On scaling contrastive representations \break for low-resource speech recognition}
\name{Lasse Borgholt$^{1, 2}$, 
      Tycho M. S. Tax, 
      Jakob D. Havtorn$^{2}$, 
      Lars Maaløe$^{2}$ and 
      Christian Igel$^{1}$}
\address{
  $^1$Department of Computer Science, University of Copenhagen, Denmark\\
  $^2$Corti, Copenhagen, Denmark\\
  borgholt@di.ku.dk}

\begin{document}
%
\maketitle

\begin{abstract}
Recent advances in self-supervised learning through contrastive training have shown that it is possible to learn a competitive speech recognition system with as little as 10 minutes of labeled data. However, these systems are computationally expensive since they require pre-training followed by fine-tuning in a large parameter space. We explore the performance of such systems without fine-tuning by training a state-of-the-art speech recognizer on the fixed representations from the computationally demanding wav2vec 2.0 framework. We find performance to decrease without fine-tuning and, in the extreme low-resource setting, wav2vec 2.0 is inferior to its predecessor. In addition, we find that wav2vec 2.0 representations live in a low dimensional subspace and that decorrelating the features of the representations can stabilize training of the automatic speech recognizer. Finally, we propose a bidirectional extension to the original wav2vec framework that consistently improves performance.

\end{abstract}

\section{Introduction}
\label{sec:intro}

\begin{figure}[!t]
  \centering
  \includegraphics[width=\linewidth]{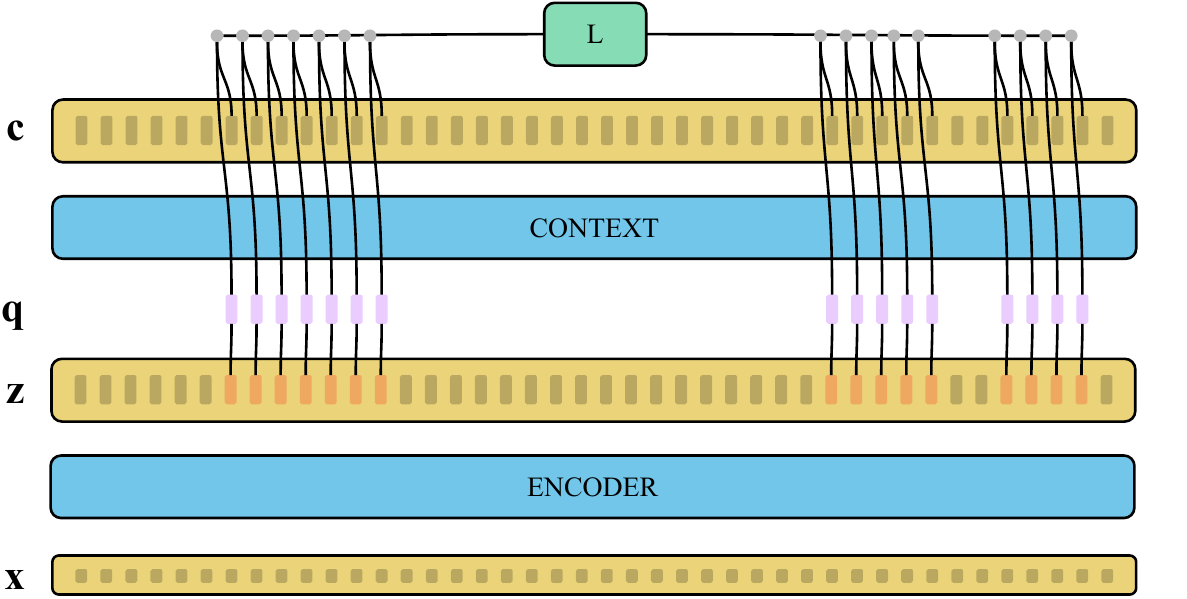}
  \caption{The wav2vec 2.0 framework \cite{baevski2020wav2vec}. The model is trained to identify the correct quantized target corresponding to the masked latent representations. The two proposed configurations have 95 and 317 million parameters respectively.}
  \label{fig:w2v2_model}
\end{figure}

Unsupervised learning for automatic speech recognition (ASR) has recently gained significant attention \cite{schneider2019wav2vec, baevski2019vq, baevski2020wav2vec, chung2019unsupervised, chung2020generative, chung2020vector, wang2020unsupervised, khurana2020convolutional, pascual2019learning}. While the majority of work has focused on learning representations encoding the input for downstream tasks \cite{schneider2019wav2vec, baevski2019vq, chung2019unsupervised, chung2020generative, chung2020vector, khurana2020convolutional, pascual2019learning}, the most promising results have been achieved with the wav2vec 2.0 framework (Fig.~\ref{fig:w2v2_model}) where a pre-trained model is fine-tuned for speech recognition. However, these models are computationally expensive due to the large amount of memory intensive transformer layers. This contradicts the promise of easily applying these representations for new ASR models on low resource languages \cite{baevski2020wav2vec}.

\begin{figure}[!t]
  \centering
  \includegraphics[width=\linewidth]{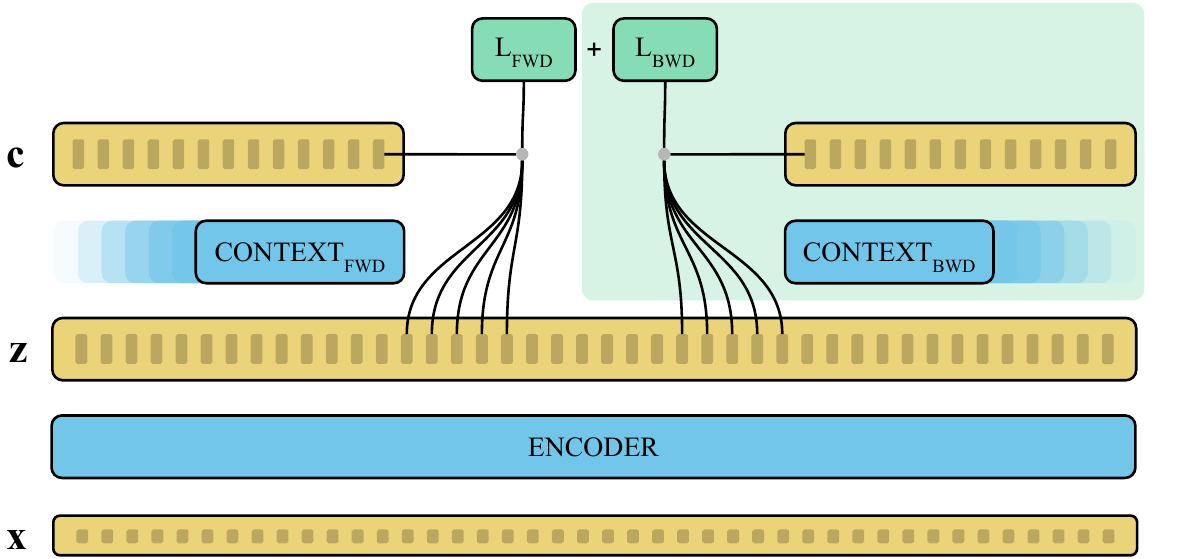}
  \caption{The wav2vec framework \cite{schneider2019wav2vec} extended with a backward context network (shaded area). The two context networks are independent, but are trained jointly with a shared encoder. The original model has 33 million parameters, while our extended model only has 18 million.}
  \label{fig:w2v1_model}
\end{figure}

In contrast to wav2vec 2.0, its predecessor (Fig.~\ref{fig:w2v1_model}) does not require fine-tuning as learned representations are used directly as input for an ASR model \cite{schneider2019wav2vec}. In addition, the pre-trained model has an order of magnitude fewer parameters than the large configuration of wav2vec 2.0. Because the frameworks are very similar, it seems obvious that representations extracted from wav2vec 2.0  would also be suitable input for training an ASR model. Training on extracted representations offers a light-weight alternative to the computationally expensive fine-tuning procedure described in \cite{baevski2020wav2vec}. 


We study how representations from the two versions of the open-source wav2vec framework compare when used as input for low-resource end-to-end speech recognition. We also propose a bidirectional extension to the original wav2vec framework that, similar to wav2vec 2.0, can use the entire latent sequence to learn contextualized representations. Our contributions are as follows:

\begin{enumerate}

\item We find that ASR models trained on the wav2vec 2.0 representations often end up in poor local minima. Decorrelating the feature space dimensions with PCA alleviates the training issues.
\item We provide an overview of ASR models trained on contrastive representations from publicly available models. Despite using a strong ASR model, performance is heavily degraded compared to fine-tuning for wav2vec 2.0. When given only 10 minutes of training data, the original wav2vec model outperforms wav2vec 2.0.

\item We propose a bidirectional extension to the original wav2vec framework. Bidirectionality consistently improves performance of ASR models trained on the representations compared to representations from unidirectional baseline models.


\end{enumerate}

\section{Contrastive Learning for Speech}
\label{sec:cpc}


\subsection{wav2vec}\label{ssec:w2v1}
In the wav2vec framework, a PCM signal $\mathbf{x} \in \mathbb{R}^T$ of sequence length $T$ is mapped to a sequence of latent representations $\mathbf{z} = \textsc{encode}(\mathbf{x}) \in \mathbb{R}^{U \times D}$ where $U$ is the downsampled sequence length ($U=T/160$) and $D$ is the dimensionality of the latent representation. This latent representation depends only locally on $\mathbf{x}$ with $\textsc{encode}(\cdot)$ parameterized by a convolutional neural network. The latent sequence $\mathbf{z}$ is fed to a context network to produce the representations used as input features for the downstream task $\mathbf{c} = \textsc{context}(\mathbf{z}) \in \mathbb{R}^{U \times D}$. In wav2vec, the context network is also convolutional but recurrent neural networks are an obvious alternative.
The model is trained with a contrastive loss function inspired by \textit{contrastive predictive coding} \cite{oord2018representation} that maximizes the similarity between a contextualized representation $\mathbf{c}_u$ and the $k$'th future latent representation $\mathbf{z}_{u + k}$ through a learned step-specific affine transformation $\mathbf{H}_k \in \mathbb{R}^{D \times D}$:
\begin{align}
    &\textsc{sim}_k(\mathbf{z}_i, \mathbf{c}_j) = \log(\sigma(\mathbf{z}_{i}^{\intercal}\mathbf{H}_k\mathbf{c}_j)) \label{eq1} \\
    &L_k(\mathbf{z}, \mathbf{c}) = -\sum_{i = 1}^{U - k} (\textsc{sim}_k(\mathbf{z}_{i+k}, \mathbf{c}_i) \sum_{d \in \mathcal{D}}\textsc{sim}_k(-\mathbf{z}_d, \mathbf{c}_i)) \label{eq2}
\end{align}

\noindent where $\sigma(\cdot)$ denotes the standard logistic function and $\mathcal{D}$ is a set of randomly sampled integers $d \sim \mathcal{U}\{1, U\}$ for indexing distractor samples $\mathbf{z}_d$. The total loss is defined as the sum over all $K$ temporal offsets, $L = \sum_{k=1}^{K}L_k$. For further details on the wav2vec architecture and training procedure see \cite{schneider2019wav2vec}.

\subsection{wav2vec 2.0}
\label{ssec:w2v2}

Similar to the first version, wav2vec 2.0 also employs an encoder and a context network, but uses a coarser downsampling ($U=T/320$) in the encoder and a transformer-based context network \cite{Vaswani2017}. In addition, a quantization network is used to learn a latent target sequence $\mathbf{q} = \textsc{quantize}(\mathbf{z}) \in \mathbb{R}^{U \times D}$. Before feeding $\mathbf{z}$ to the context network, approximately half of the $U$ time steps are \emph{masked} (i.e., replaced) by a learned $D$-dimensional vector. Given a context representation $\mathbf{c}_u$ corresponding to a masked latent vector $\mathbf{z}_u$, the model is trained to distinguish the quantized target $\mathbf{q}_u$ from distractors $\mathbf{q}_d$ sampled uniformly from the other masked time steps. The set of distractor indices $\mathcal{D}$ includes the target index $u$:
\begin{align}
    &\textsc{sim}(\mathbf{q}_i, \mathbf{c}_j) = \frac{\mathbf{q}_i^{\intercal}\mathbf{c}_j}{\norm{\mathbf{q}_i}\norm{\mathbf{c}_j}} \label{eq3}\\
    &L_u(\mathbf{q}, \mathbf{c}_u) = - \log \frac{e^{\textsc{sim}(\mathbf{q}_u, \mathbf{c}_u) / \kappa}}{\sum_{d \in \mathcal{D}} e^{\textsc{sim}(\mathbf{q}_d, \mathbf{c}_u) / \kappa}} \label{eq4}
\end{align}

\noindent where $\kappa$ is a constant temperature. The total loss is obtained by summing over all masked time steps. The model is also trained with an entropy-based diversity loss that encourages equal use of the quantized representations. The masking procedure allows for a context network consisting of multiple transformer layers that incorporate information from the full sequence instead of only time steps prior to $u$. Thus, the masking feature is key in order to be able to use an architecture well suited for fine-tuning.

\section{Bidirectional extension}
\label{sec:bde}
The context network of the original wav2vec only uses information prior to the offset latent vector $\mathbf{z}_{i+k}$. This avoids collapsing to a trivial solution and allows for online processing of streaming data. In contrast, wav2vec 2.0 requires the complete sequence at once as input to the transformer-based context network.
If we consider this setting where online processing is not required, 
the original wav2vec model can be extended with an additional context network 
that operates backward from time step $U$ to $1$. To train the backward network, the loss in equation \ref{eq2} is adapted by replacing $\mathbf{z}_{i + k}$ with $\mathbf{z}_{i - k}$. The total loss is obtained by summing the loss for the two context networks as illustrated in Fig.~\ref{fig:w2v1_model}. The context networks are independent, but trained jointly with the same encoder. The representations used for downstream tasks are the concatenation of the output from the two context networks.

\begin{table*}[t]
\begin{center}
\begin{tabular}{ l c c c c c c | c c c c} 
\toprule

 & \multicolumn{2}{c}{\textbf{10 min.}} & \multicolumn{2}{c}{\textbf{1 hour}} & \multicolumn{2}{c}{\textbf{10 hour}} & & &\\
\textbf{Name} & \textbf{clean} & \textbf{other} & \textbf{clean} & \textbf{other} & \textbf{clean} & \textbf{other} & \textbf{PCA} & \textbf{size} $D$ & \textbf{params} & \textbf{GPU days} \\
\midrule
Log mel-spectrogram & 99.6 & 99.7 & 66.5 & 82.0 & 33.8 & 57.5 & No & 80 & - & - \\
\midrule
wav2vec \cite{schneider2019wav2vec} & 71.7 & 82.5 & 43.1 & 61.9 & 24.0 & 45.8 & No & 512 & 33M & ?\footnotemark[2] \\
\midrule
wav2vec 2.0 \cite{baevski2020wav2vec} \\
\footnotesize\hspace{3mm}BASE & 79.5 & 87.5 & 38.7 & \textbf{53.6} & \textbf{14.9} & \textbf{28.4} & Yes & 768 & 95M & 102.4 \\
\footnotesize\hspace{3mm}LARGE & 91.0 & 96.1 & 50.2 & 66.5 & 20.5 & 37.5 & Yes & 1024 & 317M & 294.4 \\
\footnotesize\hspace{3mm}VOX & 94.2 & 97.2 & 42.4 & 56.8 & 16.4 & 29.8 & Yes & 1024 & 317M & 665.6 \\
\midrule
\textit{Our work:} \\
LSTM-UD-512 & 69.9 & 81.9 & 41.5 & 60.9 & 23.8 & 45.2 & No & 512 & 9.6M & 3.8 \\
LSTM-UD-2x512 & 69.2 & 81.2 & 41.0 & 61.0 & 23.2 & 44.9 & No & 1024 & 18M & 9.5 \\
LSTM-BD-2x512 & \textbf{65.2} & \textbf{77.4} & \textbf{37.9} & 56.5 & 21.0 & 42.1 & No & 1024 & 18M & 9.4 \\
\bottomrule
\end{tabular}
\end{center}
\vspace{-0.4cm}
\caption{Word error rates on the clean and other test sets of LibriSpeech for ASR models trained with representations extracted from wav2vec, wav2vec 2.0 and the authors' proposed models. Results are without an external language model. \textit{GPU days} denotes training time multiplied by the number of GPUs.} 
\label{tab:results}
\end{table*}

\section{Experiments}
\label{sec:exps}

\subsection{Data and pre-trained models}
\label{ssec:data}

The original wav2vec model was trained on the 960 hours of LibriSpeech dataset \cite{panayotov2015librispeech}. We trained our bidirectional extension and baseline models on the same data. We used three pre-trained models from wav2vec 2.0: \textsc{base}, \textsc{large} and \textsc{vox}\footnote{https://github.com/pytorch/fairseq}. The \textsc{large} model is a deeper and wider version of the \textsc{base} model. Both are trained on the 960 hour LibriSpeech. The \textsc{vox} model is identical to the \textsc{large}, but trained on 60.000 hours of speech from the LibriLight dataset \cite{kahn2020libri} which is an in-domain extension of LibriSpeech providing large quantities of unlabeled data and a standardization of smaller subsets from the original 960 hours training data. We trained ASR models on the 10 minute, 1 hour and 10 hour subsets of LibriLight for all representation models.
\footnotetext[2]{The model is trained on 16 GPUs, but training time is not stated.}

\subsection{Training procedures}
\label{ssec:models}

In addition to bidirectionality, we propose to use few filters for the first layer in the encoder network and then incrementally increase the number of filters as the temporal resolution is lowered by striding. This significantly lowers the memory footprint of the encoder by avoiding large representations while the temporal resolution is high. Thus, our encoder uses six 1D-convolutions with number of filters set to (64, 128, 192, 256, 512, 512), kernel sizes (10, 8, 4, 4, 4, 1), and strides (5, 4, 2, 2, 2, 1). With a constant filter size of 512, memory consumption would be 4.6 times higher. Each convolutional layer is followed by a group normalization layer with 32 groups \cite{wu_group_2018} and ReLU non-linearity clipped at the value of 5. Instead of using convolutions as in wav2vec, we used four LSTM layers \cite{hochreiter1997long} with 512 units each for the context network. We sampled 120 seconds of audio for each batch and trained the model for 8 epochs on LibriSpeech. We used Adam \cite{kingma2014adam} with a fixed learning rate of $3 \cdot 10^{-4}$ for the first half of training after which it was decayed to $5 \cdot 10^{-5}$. We use $K=12$ offsets and sample 10 distractors. (i.e., $\lvert\mathcal{D}\rvert = 10$).\footnote[3]{The proposed model is made available upon publication.}


For the ASR model, we used the architecture from \cite{borgholt2020context} trained with a connectionist temporal classification loss \cite{graves2006connectionist}, which has shown state-of-the-art results on the small Wall Street Journal dataset \cite{paul1992design}. The original model uses three layers of 2D-convolutions followed by 10 bidirectional LSTM layers with skip-connections and 320 units each. We replaced the 2D-convolutions with 1D-convolutions as there is no structure along the feature dimension of the learned representations. All 1D-convolutions used kernel size 3, had (640, 480, 320) units and strides (2, 1, 1). To account for the lower temporal resolution of the wav2vec 2.0 representation, strides were reduced to (1, 1, 1). We used the same optimizer and learning rate schedule as for the CPC models and batches were created by sampling up to 320 seconds of audio. The models were trained for 25k update steps on the 1 hour and 10 hour subsets, but only for 10k update steps on the 10 minute subset. Total training time was $\sim12$ hours on a single GPU for 25k updates. Results reported for the 10 minute models are averages over the 6 separate subsets of LibriLight. 

\begin{figure*}[t]
  \centering
  \includegraphics[width=\linewidth]{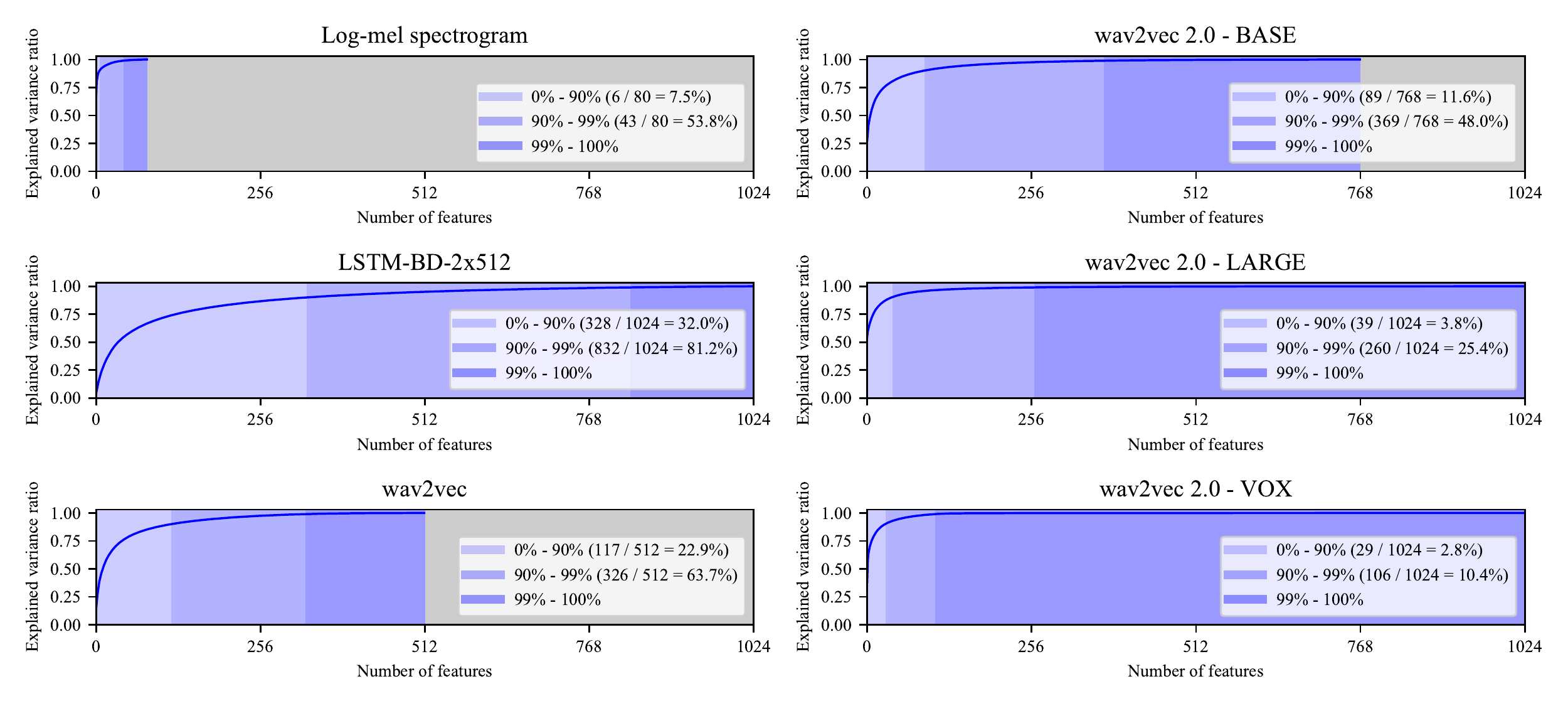}
  \vspace{-1.0cm}
  \caption{Explained variance ratio as a function of the number of features after decorrelating the feature space with PCA. The PCA transformation was computed on the 10 hour subset of LibriLight.}
  \label{fig:pca}
\end{figure*}

\section{Results}
\label{sec:results}

\subsection{Training with wav2vec 2.0 representations}
\label{ssec:w2v2_training}

We found that ASR models trained on representations extracted from the wav2vec 2.0 models had a tendency to get stuck in poor local minima. After confirming that values of the learned features followed a reasonable distribution, and that tuning the learning rate did not solve the issue, we performed a principal component analysis (PCA) of the representations. 
We found that the wav2vec 2.0 representations generally exhibited a low \emph{linear dimensionality}, that is, only few principal components are needed to explain the variance  in the representations, see Fig.~\ref{fig:pca}. Furthermore, the linear dimensionality  of the representation decreased with model complexity and the amount of training data. Indeed, the two large models were also the ones that consistently failed, while the \textsc{base} model did converge on both the 1 hour and 10 hour subsets. Feature decorrelation has previously proven useful in speech classification tasks \cite{zheng2015emotion}. Training ASR models on the decorrelated feature space, without reducing the number of features, solved the initial training issues. To ensure that the mean normalization commonly performed prior to the PCA transformation was not responsible for resolving the issue, we performed an ablation experiment where we only used mean normalization on the raw features, but this did not alleviate training issues. Representations from our models and wav2vec did not benefit from decorrelation.

\vspace{-0.3cm}
\subsection{Performance: wav2vec and wav2vec 2.0}
\label{ssec:perw2v2}
\vspace{-0.1cm}
Surprisingly, the \textsc{base} representations consistently outperformed representations from the two larger models, indicating that the quality of the learned representations does not scale with model complexity for wav2vec 2.0. For the 1 hour and 10 hour subsets, representations from the \textsc{vox} model led to better performance compared to the \textsc{large} model showing the benefits of the large increase in training data. This tendency was blurred by the poor performance of both representations on the 10 minute subset. Compared to fine-tuning results without a language model in Appendix C of \cite{baevski2020wav2vec}, performance is severely degraded despite a strong ASR model. 

Focusing on the best performing wav2vec 2.0 representations from \textsc{base}, we observe a significant word error rate reduction for the 10 hour subset compared to wav2vec. As the amount of training data is reduced, so is the difference. For the 10 minute subset, the picture is reversed as the wav2vec representations performed better.

\subsection{Performance: Bidirectionality}
\label{ssec:bd-performance}
Our baseline model (LSTM-512), as expected, yielded representations on par with the wav2vec model. The bidirectional extension (LSTM-BD-2x512) consistently improved word error rate for all subsets. To ensure the improvement is not just a result of increased model complexity, we trained another model that also used two separate context networks, but both operating in the same direction (LSTM-UD-2x512). Although this model was slightly better than the baseline model, the bidirectional model was still superior across all subsets. Furthermore, the bidirectional model gave the best result on the clean test set when trained on the 1 hour subset and for both test sets when trained on the 10 minute subset. 

\section{Conclusions}
\label{sec:cons}
We compared contrastive representations for ASR in the setting of limited training resources.
We showed that representations from the wav2vec 2.0 framework live in a low dimensional subspace. Using PCA to decorrelate the features alleviated training issues for the speech recognizer. However, ASR models trained on the fixed wav2vec 2.0 representations still performed significantly worse than the fine-tuned versions from the original wav2vec 2.0 work. Representations from the first version of wav2vec, learned with a much lower computational cost, performed better than wav2vec 2.0 on the 10 minute subset, but were inferior on the 10 hour subset. We extended the original wav2vec framework with a context network operating backwards along the temporal dimension and confirmed that bidirectionality can improve speech representations used for ASR.


\bibliographystyle{IEEEbib}
\bibliography{refs}

\end{document}